\documentclass[floatfix,preprintnumbers,superscriptaddress,showpacs,
               showkeys,prb,preprint]{revtex4} % twocolumn / preprint

\usepackage[version=3]{mhchem} % Formula subscripts using \ce{}
\usepackage{graphicx}% Include figure files
\usepackage{dcolumn}% Align table columns on decimal point
\usepackage{amsmath, amsfonts, amssymb, bm}% bold math
\usepackage{subfigure}
    
\begin{document}
\bibliographystyle{apsrev} 
\preprint{A/Fen2K6/articulo.tex}

\title{
Electronic and Geometrical Structure 
of Potassium doped
Phenanthrene
} 

\author{P.L. de Andres
\footnote{
On leave of absence from Instituto de Ciencia de Materiales de Madrid  (CSIC) 28049 Madrid (Spain)}
}

\affiliation{
Donostia International Physics Center (DIPC),
Paseo Manuel Lardizabal 4, 20018 San Sebastian, Spain.
}

\author{A. Guijarro}

\affiliation{
Departamento de Quimica Organica and Instituto Universitario de Sintesis Organica,
Universidad de Alicante, San Vicente del Raspeig, 03690 Alicante, Spain.
}

\author{J.A. Verg{\'e}s}

\affiliation{
Departamento de Teoria y Simulacion de Materiales, 
Instituto de Ciencia de Materiales de Madrid (CSIC),
Cantoblanco, 28049 Madrid, Spain.
}

\date{\today}

\begin{abstract}
The geometrical and electronic structure of potassium doped
phenanthrene, \ce{K3C14H10}, have been studied by
first-principles density functional theory.
The main effect of potassium doping is to inject charge in
the narrow phenanthrene conduction band, rendering the system
metallic.
The Fermi surface for the experimental X-rays structure
is composed of two sheets with marked one
and two dimensional character respectively. 
\end{abstract}


% 61.66.Hq Organic compounds (61.66.-f structure of specific crystalline solids)
% 61.48.-c Structure of fullerenes and related hollow and planar molecular structures
 % 81.05.ub Fullerenes and related materials in materials science.
% 71.15.Mb Density functional theory, local density approximation, gradient and other corrections
% 71.20.Rv Polymers and organic compounds.

% 74.20.Pq electronic structure calculations
% 74.70.Kn organic superconductors
% 74.70.Wz fullerenes and related materials


%Use showkeys class option if keyword display desired

\maketitle

%\section{Introduction.}

There is much interest in the properties of
polycyclic aromatic hydrocarbons (PAHs) due to their
remarkable potential in a number of fields including
electronic devices, energy storage, molecular recognition, etc. 
Phenanthrene (\ce{C14H10}) is one of the smallest molecules
of that family with very interesting electronic properties derived
partly from its "arm-chair" edge termination, as opposed to
the "zigzag" termination characteristic of the Acene family
(e.g. anthracene, with the same molecular formula).
These two different terminations have profound effects in
in the limit of very large PAHs (i.e., graphene nanoribbons) since
anthracene like strips of material are metallic while
phenanthrene like strips can be either metallic or semiconducting
depending on their widths.\cite{nakada96,yoshizawa98}
Furthermore, organic crystals based in different PAHs 
show an amazing large number of useful properties that can be
easily tuned by the addition of appropriate contaminants. 
A recent example is the discovery of a whole new family of 
organic high-Tc superconductors when doped with alkali 
metals.\cite{mitsuhashi10,wang11,coronene11,kosugi11,K6Pi2}
In particular, recent reports show that potassium doped
phenanthrene in a stoichiometry \ce{K3C14H10}
is a superconductor with $T_{C}=5$ K.\cite{wang11}
The mechanism for superconductivity is not clear yet,
but a dependence of $T_{C}$ with external pressure has
been found that hints to a non-conventional type of superconductor.
To elucidate these questions it is of paramount importance to gather information
about the geometrical and electronic structure of the material;
which is the main motivation for this work.

\begin{figure}
\includegraphics[clip,width=0.99\columnwidth]{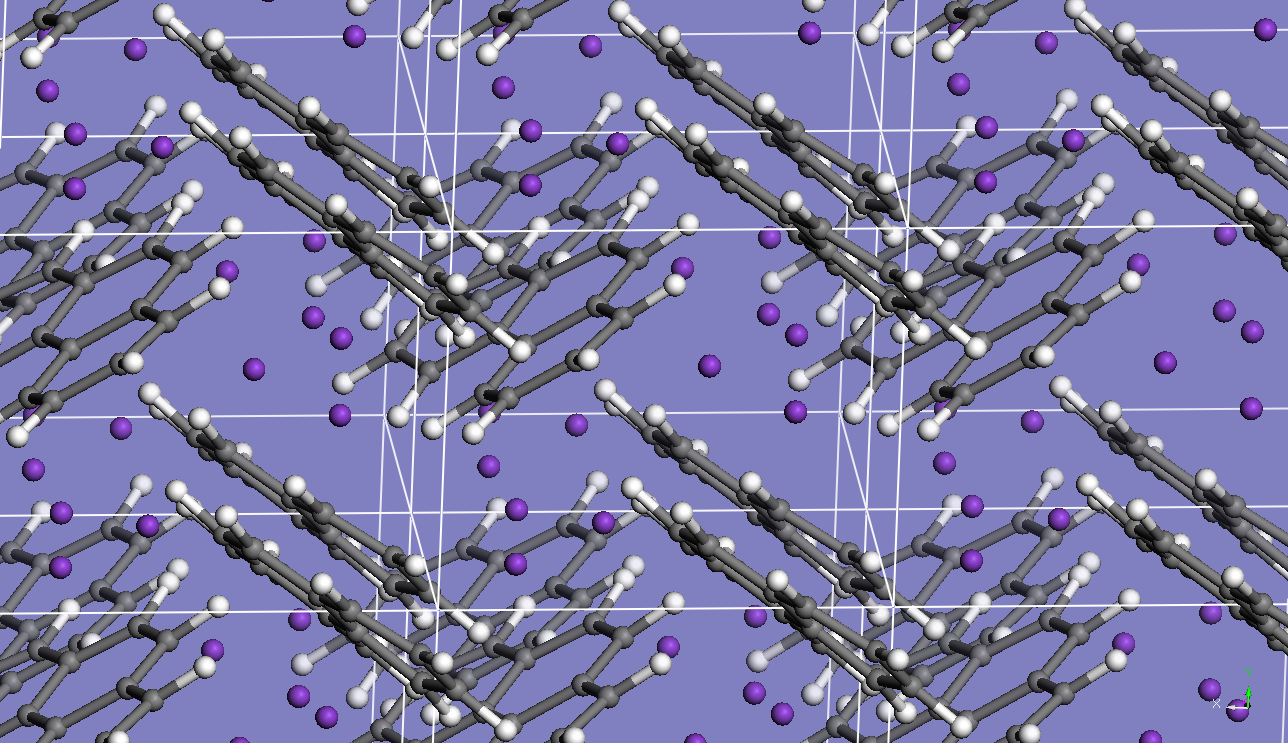} \\
\includegraphics[clip,width=0.99\columnwidth]{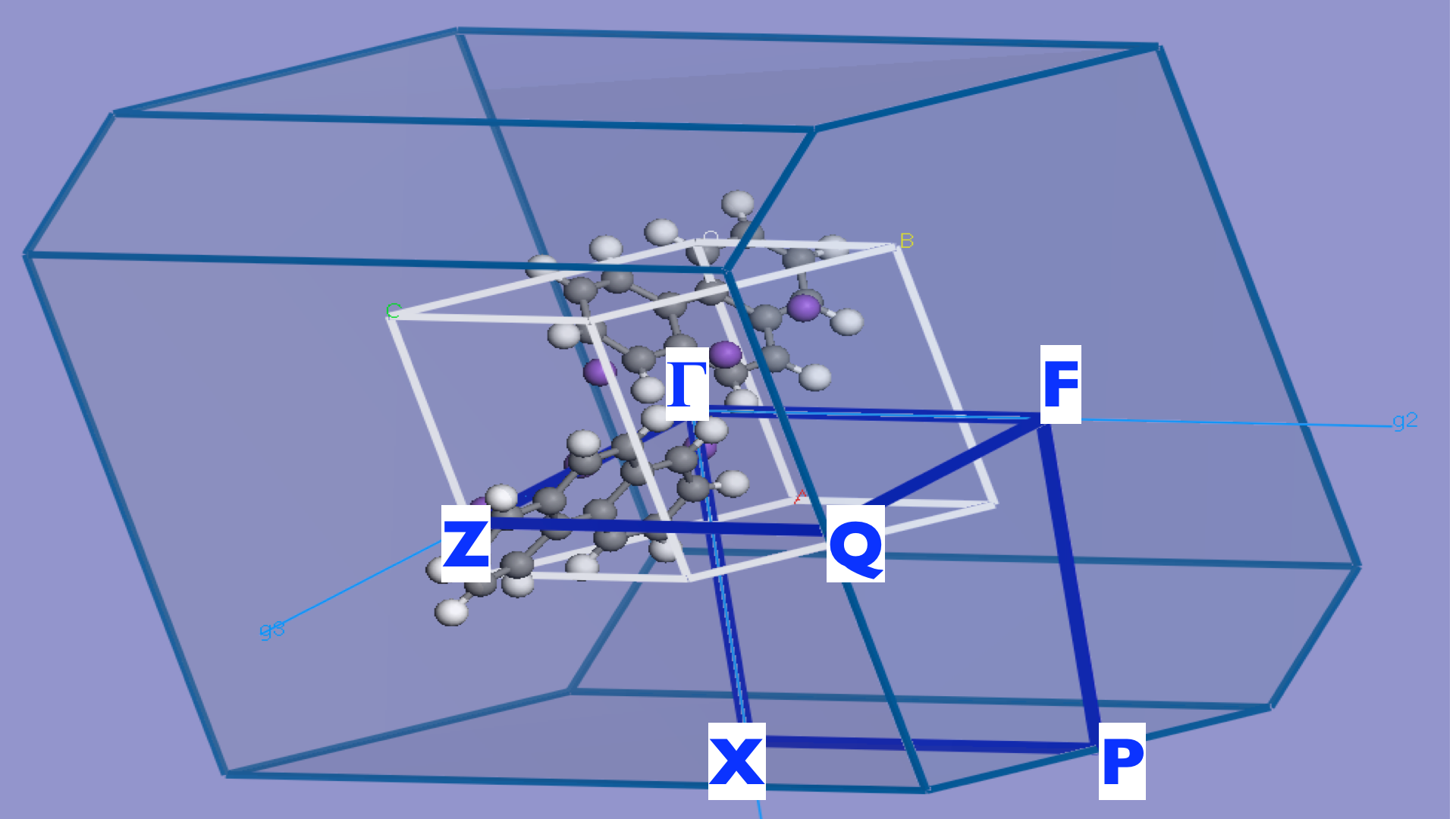}
\caption{
(a) Optimized geometry structure for \ce{K3} doped phenanthrene 
(upper pannel).
(b) The phenanthrene molecule displaying its arm-chair like
edges (lower left pannel).
(c) Brillouin zone with the selected path for the band structure
calculation shown in Fig. \ref{fgr:BSK3} (lower right pannel).
}
\label{fgr:UC}
\end{figure}

%\section{Methods.}

Theoretical calculations for model systems of interest have
been performed using ab-initio
density functional theory (DFT).\cite{Hohenberg64}
Wavefunctions have been expanded in a plane-wave basis set
up to a cutoff of $680$\,eV and were sampled on
a Monkhorst-Pack $8 \times 10 \times 7$ mesh inside the Brillouin zone.
Electronic bands were obtained using a smearing width of
$\eta=0.01$\,eV.
Carbon and hydrogen atoms have been described by 
accurate norm-conserving pseudopotentials.\cite{Lee96}
For the exchange and correlation (XC) potential the 
local density approximation (LDA) has been chosen, owning to
its well-known success dealing with these kind of systems.\cite{Kohn65}
All these choices have been proven in different occasions adequate to 
reproduce experimental geometrical parameters. 
Total energies were computed with the CASTEP
program,\cite{Clark05} as implemented in
Materials Studio.\cite{accelrys}

%\section{Results.}
%\subsection{Clean phenanthrene}
A monoclinic unit cell (UC) displaying a
P2$_{1}$ symmetry and including two
phenanthrene molecules in the basis
has been set up with
parameters derived from an X-rays analysis
(ref\cite{Trotter63}, Fig. \ref{fgr:UC}).
Using the formalism and the parameters described above 
we have optimized the system 
to minimize residual forces and stresses;
the corresponding parameters for the UC are displayed in
Table \ref{tbl:UC} and the coordinates of all the atoms
in the basis can be retrieved from the complementary
material. According to the well known
tendency of LDA to underestimate bond lengths the size of
the UC is reduced in the three crystallographic axis 
with respect to the experimental values by
$\approx 5-3$\% resulting in an overall decrease of the
unit cell volume of $11$\%.
Most of the atomic coordinates in the phenanthrene molecules forming
the unit cell basis agree quite well with the experimental ones
(i.e. $\pm 0.05$ {\AA}, we notice
that different published experimental determinations of the crystal could
differ by $\pm 0.02$ {\AA}).
The exception comes from the four H atoms participating
in the CH-$\pi$ interaction between
first-neighbors molecules; the theory to experiment difference in
these can reach $0.18$ {\AA}. 
This is not at all surprising since 
the X-rays analysis of H is itself so
difficult, and because
we are bound to compare a theoretical calculation
performed at T=0 K with an experimental structural determination performed at
room temperature where the thermal vibrations should affect more to light atoms
like hydrogen, especially if they form part of a sensitive bond keeping 
together the two molecules in the unit cell.
Finally, the average angle between the planes containing
the two molecules stays within $2^{\circ}$
of the experimental value.
Therefore, these results confirm the adequacy and set the
accuracy of the proposed formalism.

\begin{table}
\caption{ 
\label{tbl:UC}
Unit cell parameters for clean and potassium doped
phenanthrene determined either from X-rays
structural analysis or by DFT full optimization 
of forces and stresses (distances in {\AA} and angles in
degrees)
}
\begin{tabular}{|l|r|r|r|r|r|}
\hline
  & a & b & c & $\beta$ & V \\
  \hline
\ce{C14H10} & 8.05 & 5.96 & 9.16 & 96.8 & 437 \\
\ce{K3C14H10} & 8.04 & 6.54 & 10.09 & 102.6 & 517 \\
\hline
\ce{C14H10}\cite{Trotter63} & 8.46 & 6.16 & 9.47 & 97.7 & 489 \\
\ce{K3C14H10}\cite{wang11} & 8.65 & 5.96 & 9.30 & 100.2 & 472 \\
\hline
\end{tabular}
\end{table}

%\subsection{Potassium doped phenanthrene }
Experimental analysis of \ce{K3C14H10} doped crystal
shows a contraction of the unit cell in the $\vec b$ and $\vec c$ 
directions, a expansion in the $\vec a$ direction,
and a small increase in the angle $\beta$. The unit
cell volume is overall decreased by -3.5\%.
From a theoretical point of view, however, intercalation
of potassium creates an internal stress that results in
an increase of the unit cell volume by +18\%. 
Optimizing the unit cell to remove 
the stress field yields 
an energy improvement of $0.9$ eV/unit cell.
($max(S_{ij}) \le 0.1$ GPa, 
$max(F) \le 0.1$ eV/{\AA}).
At the present moment there is no a reliable 
structural determination that could yield all the positions of
the atoms in the unit cell basis, therefore our best knowledge on
the particular positions of interstitial potassium is limited to 
a procedure like the one employed here,  
that in the worst case
should represent at least a local metastable one
intervening with a given weight in the appropriate 
thermodynamical
average at a given T.

%\subsection{Electronic structrure}

\begin{figure}
\includegraphics[clip,width=0.99\columnwidth]{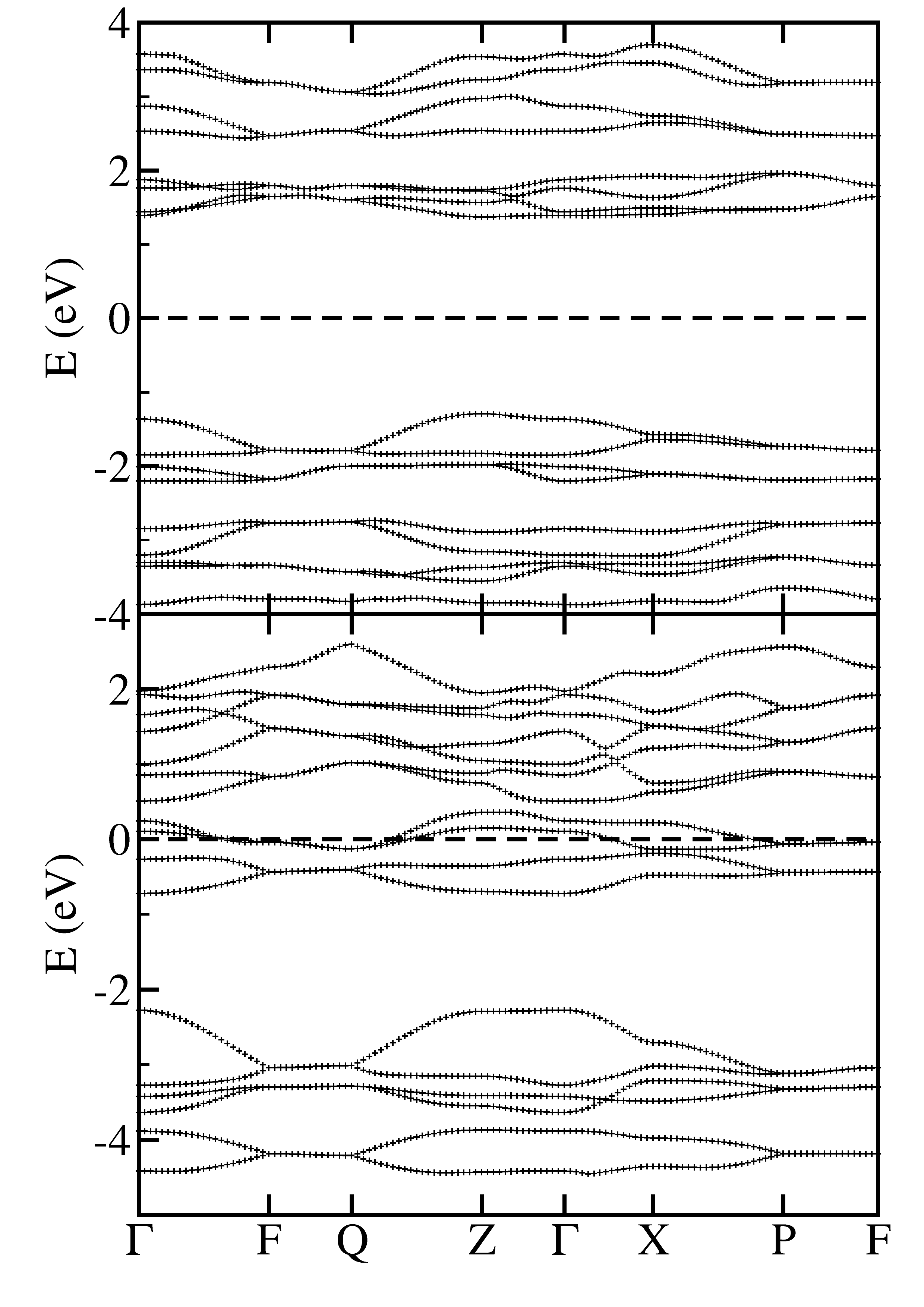} 
\caption{
Band Structure for the phenanthrene crystal (upper pannel)
and for the \ce{K3} doped crystal corresponding to the X-rays
structural determination\cite{wang11} (lower pannel).
The Fermi energy has been taken as origin for energies (dashed lines).
}
\label{fgr:BSK3}
\end{figure}

\begin{figure}
\includegraphics[clip,width=0.99\columnwidth]{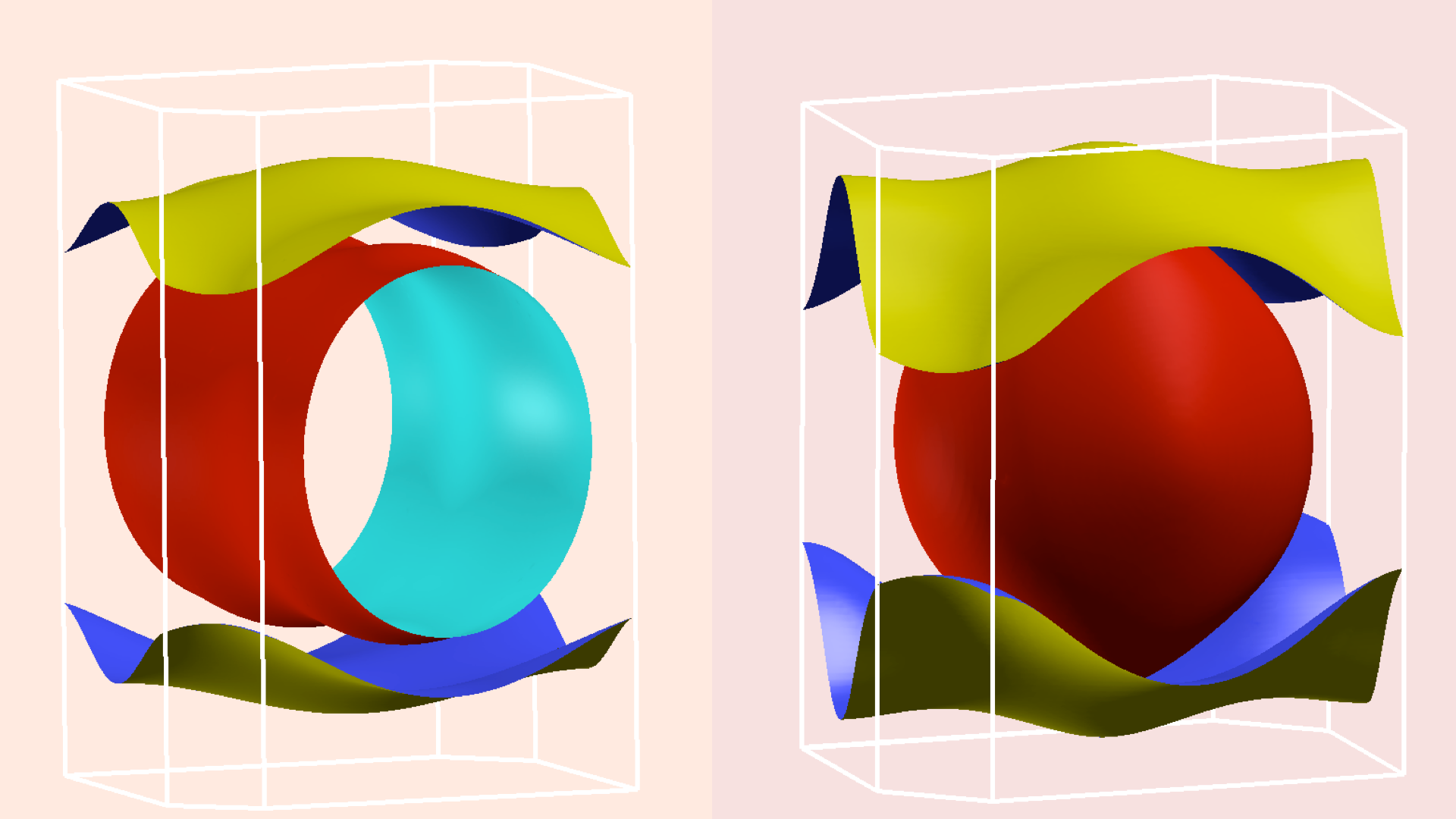} 
\caption{
For the doped phenanthrene crystal the
Fermi surface corresponding to the X-rays determination
(left pannel) is compared to the one related to the
relaxed unit cell (right pannel). 
}
\label{fgr:Fermi}
\end{figure}

We compute the electronic band structure along a selected path
on the Brillouin zone (Fig. \ref{fgr:UC}). Our LDA calculation 
yields for the phenanthrene crystal
a direct gap at $\Gamma$ of $2.75$ eV,
to be compared with an experimental one
of $3.16$ eV.\cite{yoshizawa98,bhatti00}
The important features
in the band structure come from the four bands at the top of the valence
band and the four bands at the bottom of the conduction band. 
As it has been discussed in similar systems,
these show a strong molecular character with weak overlap,
and should be related to the HOMO and HOMO-1 and LUMO and LUMO+1
molecular orbitals respectively.\cite{kosugi11}
The main effect
for the crystal intercalated with potassium, \ce{K3C14H10},
is an effective doping of the conduction band that renders the
system metallic (Fig. \ref{fgr:BSK3}).
Our calculations show that there is an important charge transfer
from the alkali atom to the organic molecules and
the Fermi energy is located in the middle of a narrow band that
becomes responsible for the metallic properties of the 
doped material.

The experimental unit cell used in these calculations, however,
is under an approximate isotropic pressure of $3$ Gpa in the
present formalism. As the $T_{C}$ of this material
seems to be sensitive to external pressure, we have checked
the effect of removing the stress by letting the unit cell relax to
an stable equilibrium condition. The unit cell increases slightly in volume,
decreasing the hybridization between the LUMO orbitals of both
molecules and the conduction band crossing the Fermi energy
becomes less dispersive, but no big changes are observed
along the chosen path to represent the bands.
The Fermi surface, however, is more sensitive to small geometrical 
details: we have compared the results for the relaxed and 
unrelaxed unit cells (Fig. \ref{fgr:Fermi}).
To improve the sampling in k-space and the representation of
surfaces in the Brillouin zone, a 4x4 tight-binding
hamiltonian describing the first four conduction bands has been
derived by fitting the electronic bands on a 
dense Monkhorst-Pack
grid in the irreducible part of 
the Brillouin zone (280 k-points). A two
sheet Fermi surface has been obtained that is
compatible with the ab-initio determination
in a coarse k-space grid. 
For the X-rays derived UC, one of the sheets is nearly planar
and hints to the strong one-dimensional character of states
involved in the metallic conduction on this material. The other
sheet is nearly cylindrical and reveals a more two-dimensional
character (Fig. \ref{fgr:Fermi}). 
After relaxing the unit cell, the planar sheet changes very little,
but the cylindrical-like shape becomes an spheroid where the
BZ boundaries are now not touched and corresponding gaps
happening near the Fermi energy are closed.

\begin{figure}
\includegraphics[clip,width=0.99\columnwidth]{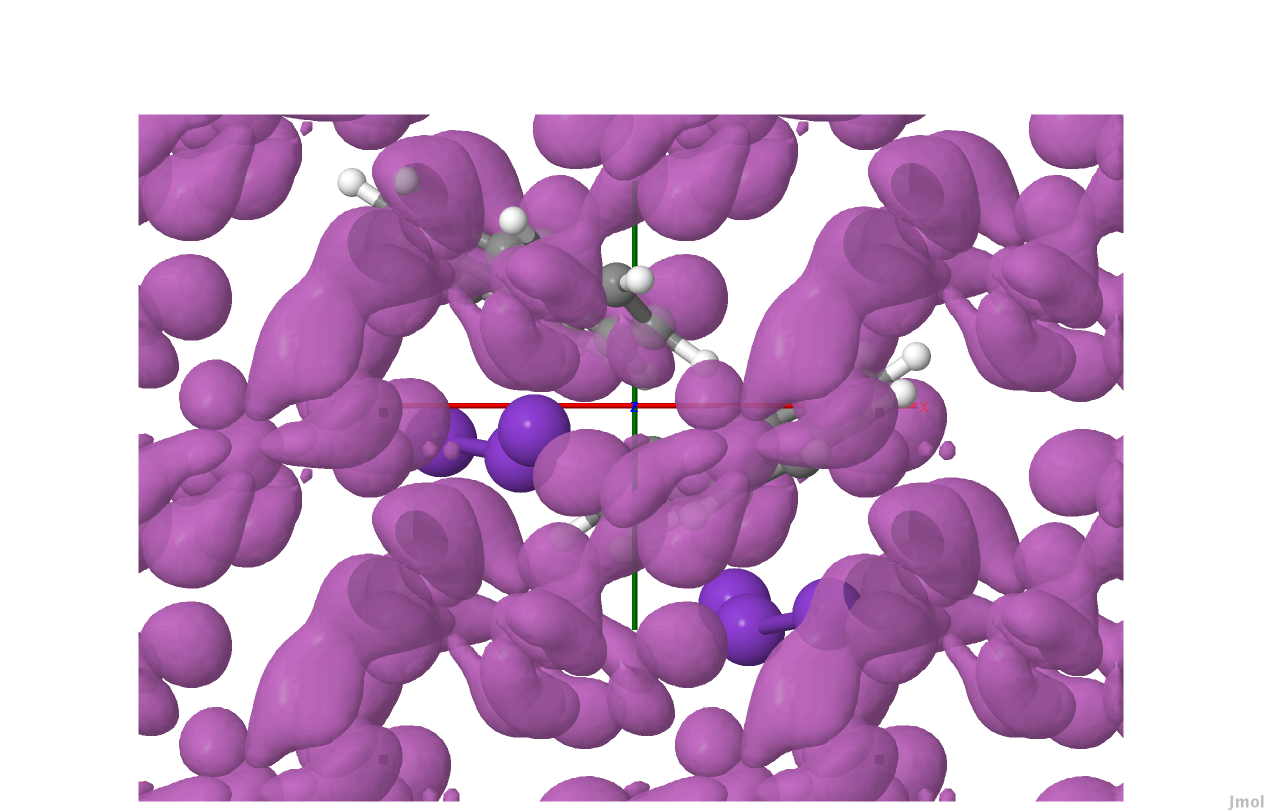}
\caption{
Iso-density corresponding to the four bands nearest the Fermi energy
inside the unrelaxed unit cell.
}
\label{fgr:isoden}
\end{figure}

The spatial distribution of electronic density confirm the above picture:
we draw in Fig. \ref{fgr:isoden} surfaces of iso-density that
integrate the four relevant bands close to the Fermi energy.
We observe the build up of extended
metallic states based in the overlap of $\pi$-like orbitals 
originated from both phenanthrene molecules. 
This is consistent with our image of an important charge transfer from
potassium to phenanthrene to build the metallic phase.

%\section{Conclusions.}

We have studied the geometrical and electronic structure of the clean and
potassium doped phenanthrene crystal. 
The main effect of doping is
the population of the phenanthrene conduction band making the system
metallic. 
According to our calculations the new metallic state is basically related
to the properties of  molecular orbitals
The experimental determination of the UC by X-rays diffraction
techniques implies stress since the UC volume is predicted to shrink by
$\approx 10$\% with respect to the pristine crystal. 
The Fermi surface under stress shows a low-dimensional character
that can be transformed by allowing the UC to deform to a global
equilibrium shape.

%\section{Acknowledgments.}
This work has been financed by the Spanish
MICINN (MAT2008-1497, CTQ2007-65218, CSD2007-6,
and FIS2009-08744),
DGUI of the Comunidad de Madrid (MODELICO-CM/S2009ESP-1691)
and MEC (CSD2007-41, "NANOSELECT").

% \bibliography{1articulo} % Produces the bibliography via BibTeX.
% include *.bbl

%\clearpage
%\newpage

\section{APPENDIX: additional material}

\begin{table}
\caption{\label{tbl:X}
Fractional coordinates
for the atoms forming the basis in the 
potassium-doped fully relaxed structure
(the corresponding UC has been defined
in Table \ref{tbl:UC}).
}
\begin{tabular}{crrr}
\hline
ATOM & u & v & w \\
\hline
H & 0.2801& 0.1603&-0.3413\\
H & 0.0165&-0.0229&-0.4506\\
H &-0.1379&-0.2386&-0.3076\\
H &-0.0311&-0.2629&-0.0625\\
H & 0.0314&-0.2411& 0.1399\\
H & 0.1042&-0.1898& 0.3868\\
H & 0.3438& 0.0317& 0.4950\\
H & 0.5360& 0.1894& 0.3535\\
H & 0.5755& 0.2884& 0.1222\\
H & 0.4811& 0.2857&-0.1280\\
H &-0.2802& 0.6623& 0.3408\\
H &-0.0161& 0.4803& 0.4506\\
H & 0.1382& 0.2634& 0.3081\\
H & 0.0312& 0.2379& 0.0634\\
H &-0.0313& 0.2580&-0.1393\\
H &-0.1038& 0.3085&-0.3861\\
H &-0.3431& 0.5293&-0.4949\\
H &-0.5364& 0.6875&-0.3540\\
H &-0.5761& 0.7881&-0.1230\\
H &-0.4818& 0.7863& 0.1272\\
C & 0.2110& 0.0720&-0.2768\\
C & 0.0583&-0.0287&-0.3395\\
C &-0.0288&-0.1423&-0.2604\\
C & 0.0377&-0.1604&-0.1194\\
C & 0.1408&-0.1409& 0.1823\\
C & 0.1825&-0.1139& 0.3247\\
C & 0.3241& 0.0000& 0.3861\\
C & 0.4273& 0.0913& 0.3047\\
C & 0.4708& 0.1858& 0.0760\\
C & 0.4162& 0.1845&-0.0683\\
C & 0.2711& 0.0770&-0.1336\\
C & 0.1792&-0.0407&-0.0495\\
C & 0.2314&-0.0357& 0.0954\\
C & 0.3806& 0.0843& 0.1601\\
C &-0.2114& 0.5731& 0.2765\\
C &-0.0582& 0.4735& 0.3395\\
C & 0.0291& 0.3596& 0.2607\\
C &-0.0378& 0.3405& 0.1199\\
C &-0.1408& 0.3580&-0.1818\\
C &-0.1822& 0.3848&-0.3243\\
C &-0.3240& 0.4980&-0.3860\\
C &-0.4281& 0.5888&-0.3051\\
C &-0.4704& 0.6867&-0.0766\\
C &-0.4162& 0.6856& 0.0678\\
C &-0.2714& 0.5778& 0.1333\\
C &-0.1796& 0.4595& 0.0496\\
C &-0.2318& 0.4634&-0.0953\\
C &-0.3808& 0.5835&-0.1604\\
K & 0.1882&-0.4609&-0.3002\\
K & 0.2300&-0.4875& 0.0179\\
K & 0.4021&-0.4436& 0.3237\\
K &-0.1877& 0.0423& 0.3030\\
K &-0.2305& 0.0102&-0.0160\\
K &-0.4018& 0.0531&-0.3221\\
\hline
\end{tabular}
\end{table}

\end{document}